\documentclass[twocolumn,preprintnumbers,amsmath,amssymb]{revtex4}

\usepackage{dcolumn}
\usepackage{amstext}
\usepackage{amsmath}
\usepackage{amssymb}
\usepackage[pdftex]{graphicx}
\usepackage{subfigure}
\usepackage[T1,OT1]{fontenc} 

\begin{document}
\title{Fluid membranes can drive linear aggregation of adsorbed spherical nanoparticles}
\author{An{\fontencoding{T1}\selectfont\dj}ela \v{S}ari\'c and Angelo Cacciuto*}
\affiliation{Department of Chemistry, Columbia University\\ 3000 Broadway, MC 3123\\New York, NY 10027 }
\renewcommand{\today} 

\begin{abstract}
Using computer simulations we show that lipid membranes can mediate linear aggregation of spherical nanoparticles binding to it for a wide range of biologically relevant bending rigidities. This result is in net contrast with the isotropic aggregation of nanoparticles on fluid interfaces or the expected clustering of isotropic insertions in biological membranes. We present a phase diagram indicating where linear aggregation is expected, and compute explicitly the free energy barriers associated with linear and isotropic aggregation. Finally, we provide simple scaling arguments to explain this phenomenology.
\end{abstract}
\maketitle

Lipid membranes have unique mechanical properties that are crucial for many biological processes, including cellular recognition, signal transduction, inter- and intracellular transport, and cell adhesion. Most of these processes require interactions of a lipid-bilayer with a variety of nano- and micro-size objects, such as proteins, DNA, viruses and other biomacromolecules. Along with its fundamental importance, understanding the interactions of fluid membranes with nano-objects is a crucial component in targeted drug-delivery design and in nanotoxicity studies. It also has intriguing implications for medical imaging~\cite{kostarelos} and for the development of biosensors and functional biomimetic materials~\cite{costanzo,webb}.

Lipid membranes are typically very flexible and
under thermal perturbations they undergo surface deformations that are significantly larger than their thickness. Because of such a flexibility, they can easily be deformed when interacting with nano-particles that can be either adsorbed on the membrane surface or embedded in the lipid bilayer. The resulting membrane deformations may in turn  mediate interactions between the membrane-bound objects. This phenomenon has been extensively studied over the past two decades, both experimentally and theoretically. Most of the previous studies focused on the interactions between embedded inclusions. A bending-mediated Casimir-like isotropic interaction was initially proposed as a possible mean of driving protein aggregation on a lipid bilayer~\cite{safran, pincus}. Later papers have shown that a more accurate accounting of the local constraints imposed by non-isotropic inclusions on the membrane can lead to additional complex terms whose sign and functional form is very much dependent on how the objects anchor to the surface; see for instance ~\cite{netz,fournier,chou,deserno,chen} and references therein.
Hydrophobic mismatch~\cite{killian}, difference in curvature between the membrane and the embedded objects~\cite{leibler,lipowsky,pep} or line tension between the lipids and the inclusions~\cite{baumgart} can also induce domain formations. 

Adsorption or inclusion of objects comparable in size to the membrane thickness ($\sim$ 5nm) greatly perturbs the local packing of the lipids leading to quite complex phenomena dependent on the molecular details of the membrane-object interactions.
When considering larger objects, on the contrary,  it becomes feasible to describe the membrane as a continuous surface
and coarse-grain its interactions with the nanoscopic objects with generic binding potentials.
Here we are interested in membrane driven interaction between adsorbing colloidal particles that are 
more than one order of magnitude larger than the membrane thickness.
Despite their structural complexity, for sufficiently large scales the behavior of lipid membranes  can be described by a small number of elastic parameters that capture their response to deformation; a bending rigidity $\kappa_b$ of the order of $10k_{\rm B}T$, and a small surface tension $\gamma\approx 10^{-2}-10^{-3} pN/nm$ are the most important ones. Both can be altered either by dispersing within the bilayer additional molecular components, or by changing the lateral forces/osmotic pressure applied on the membrane.

In this paper we show that spherical nanoparticles adhering to fluid membranes can self-assemble into a variety of two-dimensional aggregates.
Significantly, for intermediate and biologically relevant values of the bending rigidity we find that particles preferentially arrange into
linear/flexible aggregates. This result is in striking contrast with most of the theoretical studies on membrane inclusions that 
predict isotropic aggregation when the embedding object imposes an isotropic deformation on the surface. 
Linear aggregation is expected only for sufficiently anisotropic  wedge-like local deformations~\cite{fournier}, and this is clearly not 
the case for spherical nanoparticles. We find that the key to understand the stability of linear versus isotropic aggregates resides in the interplay between bending and binding energies of the nanoparticles.
The latter  term, usually and correctly neglected when dealing with embedded nano-components, does indeed play a major role in the structural morphology of the aggregates  formed by non-embedded adhering components.

It should be stressed that string-like formations very similar to those we present here have been observed experimentally in several systems.
For instance, colloidal particles bound to giant phospholipid vesicles (GUV) via streptividin-biotin bonds or by electrostatic physisorption form one-dimensional ring-like assembles~\cite{safinya}. Similarly, the cationic lipid-DNA complexes of low net charge assemble into linear colloidal aggregates when adsorbed to the cell membrane~\cite{safinya1}. However, to the best of our knowledge, this phenomenon has never been explained, nor studied in detail. In this paper we use a combination of numerical simulations and scaling arguments to detail the physical origin behind it.

We performed Monte Carlo simulations of planar and spherical fluid membranes interacting with adsorbing nanoparticles.
The membrane is modeled using a simple one particle-thick solvent-free model, and consists of $N$ hard spherical beads, of diameter $\sigma$, connected by flexible links to form a triangulated network~\cite{ho,nelson,gompper} whose connectivity is dynamically 
rearranged to simulate the fluidity of the membrane. 
The membrane bending energy acts on neighboring triangles, and has the form 
\begin{equation}
E_{ij}=\frac{\kappa_b}{2}(1-{\bf n}_i\cdot {\bf n}_j)\,,
\end{equation}
where $\kappa_b$ is the bending rigidity, and the ${\bf n}_i$ and ${\bf n}_j$ are the normals of two triangles $i$ and $j$ sharing a common edge.  The cost associated with area changes is included via the energy term 
 $E_{\gamma}=\gamma A$, where $\gamma$ is the tension of the surface and $A$ is its total area.
 A nanoparticle is modeled as a sphere of diameter $\sigma_{np}=Z\sigma$, with $Z=3,4$ or $6$. Excluded volume between any two spheres in the system (nanoparticles and surface beads) is enforced with a hard-sphere potential. 
Finally, the nanoparticle-to-surface adhesion is modeled via
a generic power-law potential between the nanoparticles and the surface beads defined as
\begin{equation}
V_{\rm {att}}(r)=-D_0 \left(\frac{\sigma_M}{r}\right)^{6}
\end{equation} 
with $\sigma_M=(\sigma+\sigma_{np})/2$, and cutoff at $r_{\rm cut}=1.5\sigma_M$.  
Following~\cite{smit}, simulations of the planar membrane were carried out in the $N\gamma T$ 
ensemble, while the $NVT$ ensemble was used for the spherical membrane. 
In each simulation the number of nanoparticles  is held constant,
and the surface tension is set to $\gamma=3 k_{\rm{B}}T/\sigma^2$. 
For $\sigma\approx 30-50nm$ we have nanoparticles of diameter $\sigma_{np}=100-200nm$ and surface tensions 
$\gamma\approx10^{-2}-10^{-3} pN/nm$.

\begin{figure*}
\includegraphics[width=100mm]{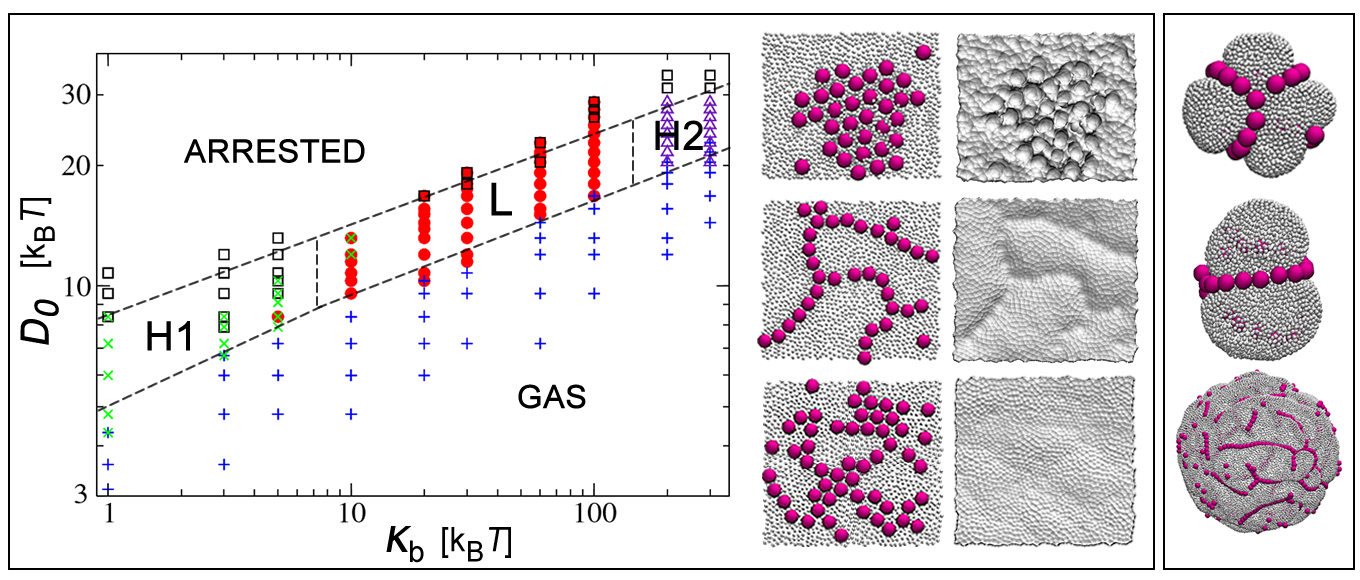}
\caption{Left panel: Phase diagram of particle self-assembly on a fluid surface in terms of the surface bending rigidity $\kappa_b$ and particle binding energy $D_0$. The snapshots show typical aggregates in the {\bf H1}, {\bf L} and {\bf H2} phases in a top-to-bottom order, and the deformation pattern they leave on the membrane.  The membrane-area is $A\simeq(40 \times 40)\sigma^2$, the nanoparticle-size $\sigma_{np}=4\sigma$ and their surface fraction $\rho=0.27$. Right panel: Snapshots of the linear aggregates on the spherical membrane. The upper two snapshots show the system of $R\simeq 15\sigma$, $\sigma_{np}=4\sigma$ and $\rho=0.11$ and the bottom snapshot depicts $R\simeq 45\sigma$, $\sigma_{np}=3\sigma$ and $\rho=0.16$.}
\label{fig1} 
\end{figure*}

We begin by computing the phase behavior of the system for different values of the surface bending rigidity $\kappa_b$ and nanoparticles' adhesive energy $D_0$. The results are summarized in the left panel of Fig.~\ref{fig1}, and  report the structure of the aggregates observed for each pair $[\kappa_b,D_0]$ in the case of the planar geometry.
A gas phase occurs when $D_0$ is too weak for the particles to deform the membrane. 
In this phase particles are just lightly bound to the surface, they are highly mobile, and have a certain probability of detaching from it. 
An arrested phase occurs for  large values of $D_0$. In this case particles bind very strongly to the membrane resulting 
in large local deformations that heavily limit their mobility over the surface. This typically leads  to configurations 
that are kinetically trapped  or even to nanoparticle engulfment.
Three ordered phases occur for moderate values of $D_0$. 
Each of the three phases spans a range of $\kappa_b$ values. 
For small values of the bending rigidity,  particles create well defined deep-spherical imprints in the membrane and organize into ordered hexagonal arrays (\textbf{H1}). Low cost in bending energy and high gain in surface binding allows for these deep deformations. In this phase the nanoparticles are not in direct contact with each other, but are separated by the pinched parts of the membrane. 
Close-packing maximizes sharing of the pinched regions between neighboring nanoparticles, thus maximizing the surface-to-nanoparticle contact area. 
An identical result is obtained when repeating the simulations on the spherical membrane, and is reminiscent of the experimentally observed two-dimensional hexagonal crystal formed by
negatively charged particles on positively-charged surfactant vesicles reported in~\cite{weitz}.
Even in this case the colloids are extensively wrapped by the membrane and are not in direct contact with each other. 

For biologically relevant values of $\kappa_b$, our nanoparticles create smooth channel-like distortions on the membrane and self-assemble into linear aggregates (\textbf{L}) non unlike those predicted for anisotropic membrane inclusions~\cite{fournier}.
Although we have not computed a structural phase diagram for our vesicle model, we find that  the simulations on the vesicle performed at
different nanoparticle concentrations and vesicle radii lead to analogous results.
Here particles form sinuous linear patterns that tend to follow the equatorial lines of the vesicle. Snapshots from our simulations are shown in the right panel of Fig.~\ref{fig1}. This phase strikingly resembles the linear aggregates of colloidal particles on Giant Phospholipid Vesicles (GUV) obtained in~\cite{safinya}. 

For very large values of $\kappa_b$ the nanoparticles re-organize into the familiar hexagonal lattice, however, unlike what happens
for the small $\kappa_b$ aggregates, the membrane now remains almost completely flat and the nanoparticles are in contact with each other (\textbf{H2}). Because of its high stiffness, particles can only weekly deform the membrane to gain in binding energy, as a result the binding energy 
is minimized by recruiting the largest number of membrane beads in the vicinity of the nanoparticles. This effectively drives the crystallization of  
the region of the membrane that directly interacts with the nanoparticles, creating a line tension between crystalline and fluid 
membrane regions that is minimized when isotropic aggregation takes place~\cite{lipowsky,pep}.

  
As mentioned before, the formation of linear aggregates is quite surprising.
To ensure that our results are not affected by the triangulation underlying the definition of our membrane model, 
we repeated our simulations using the coarse-grained, but tether-free model proposed by Zhang et al.~\cite{li}. This model also
accounts for possible topological changes in the surface, however the elastic properties of the membrane are not fed to 
the system in the form of parameters of an elastic energy, but are  encoded into the molecular details of the anisotropic pair potentials
between the effective building blocks of the membrane, and need to be extracted by analyzing the fluctuations spectrum of the surface~\cite{li}, or by other means.
It is comforting to report that no qualitative difference was found on the overall phenomenology of the phase diagram: linear aggregates do indeed form and are not an artifact of our model. We also checked that linear aggregates do no form when limiting the
area of the particles' binding region to enforce a finite (constant) contact angle between particles and membrane. 
This case is basically equivalent to enforcing isotropic regions with intrinsic curvature, mimicking for instance the local perturbation of a protein,
in a lipid bilayer for which isotropic aggregation is expected~\cite{pep}.

To understand why linear aggregates become more favorable for moderate bending rigidities,
we placed three nanoparticles, $A, B$ and $C$ in linear formation and at a kissing distance over a planar membrane, and calculated the free energy cost required to disrupt the linear arrangement. 
The idea is to keep in place particles $A$ and $B$ and force particle $C$ to form an angle $\varphi_0$ between the vector connecting particles $A$ and  $B$ and that connecting particles $B$ and $C$  while keeping the relative particle distance unaltered.
\begin{figure}
\center
\subfigure[]{\label{fig2a}\includegraphics[width=42mm]{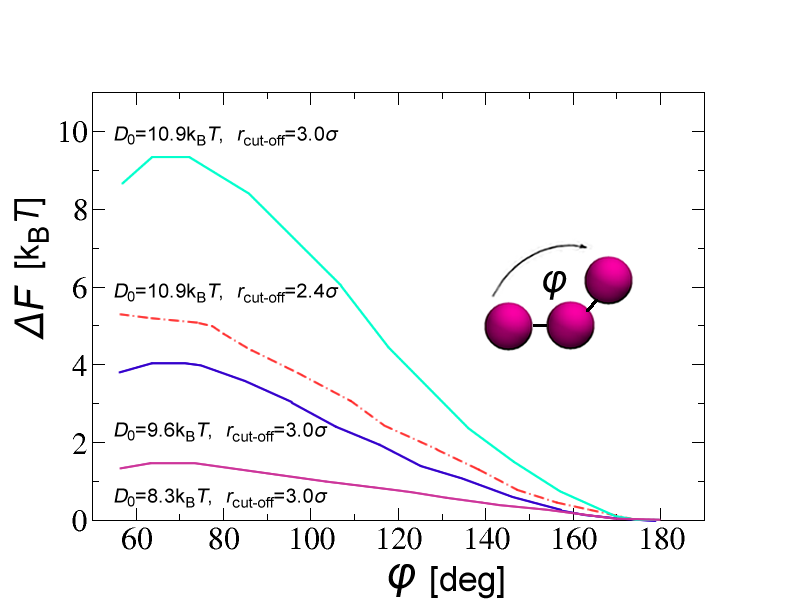}}
\subfigure[]{\label{fig2b}\includegraphics[width=42mm]{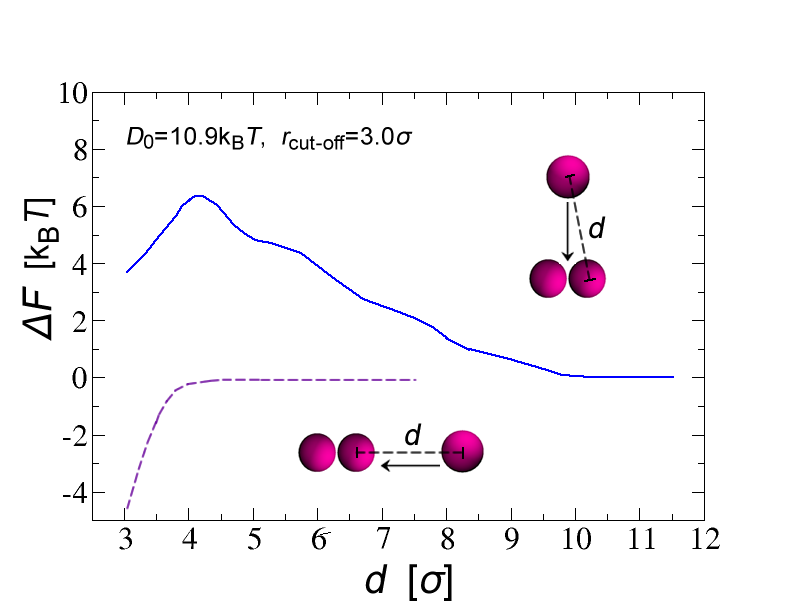}}
\caption{(a) Angular free-energy profile for three nanoparticles bound to the membrane at different values of binding constant $D_0$ and interaction range $r_{\rm{cut}}$. (b) Free-energy as a function of the separation when a third particle approaches a fixed dimer
along the direction of the dimer's axis (dashed line) and perpendicular to it (full line). 
In both cases we used $\kappa_b=20k_{\rm{B}}T$ and $\sigma_{np}=3\sigma$.}
\end{figure}
Using the umbrella sampling method~\cite{valleau}, we can 
reconstruct piece-wise the probability that the trimer arranges according to any of the explored angles, which in turns gives us access to the free energy  difference $\Delta F=F(\varphi)-F(\pi)$. All simulations were repeated for different values of $D_0$ and two different ranges of the binding potential. The results are shown in  Fig.~\ref{fig2a}, and  undoubtedly tells us that in this region of the phase diagram the linear configuration is the most stable one, with the close-packed compatible configuration ($\varphi_0=\pi/3$) sitting in a metastable shallow minimum of the free-energy curve  separated from the linear configuration ($\varphi_0=\pi$) by a significant barrier. The height of the barrier depends on the exact parameters, but is typically larger than 4$k_{\rm{B}}T$ inside the linear region of the phase diagram. 
Fig.~\ref{fig2b} shows  the free-energy as a function of particle separation when the third particle approaches the other two from infinity, either in the linear or perpendicular alignment, as depicted in the insets in Fig.~\ref{fig2b}. When the third particle approaches the dimer to form a linear aggregate,  the free-energy (when particles are sufficiently close) decreases monotonically down to a minimum at contact. When the third particle approaches the dimer from a direction that is perpendicular to the dimer's axis, we observe a repulsive free energy barrier that precedes a shallow minimum at contact. Remarkably the range of the repulsion is felt as far as three
nanoparticle diameters (up to 9 times the range of the attractive part), 
revealing correlations in the three body interactions that are significantly longer than the ones expect from a simple Casimir effect~\cite{netz,fournier}.

\begin{figure}
\center
\includegraphics[width=80mm]{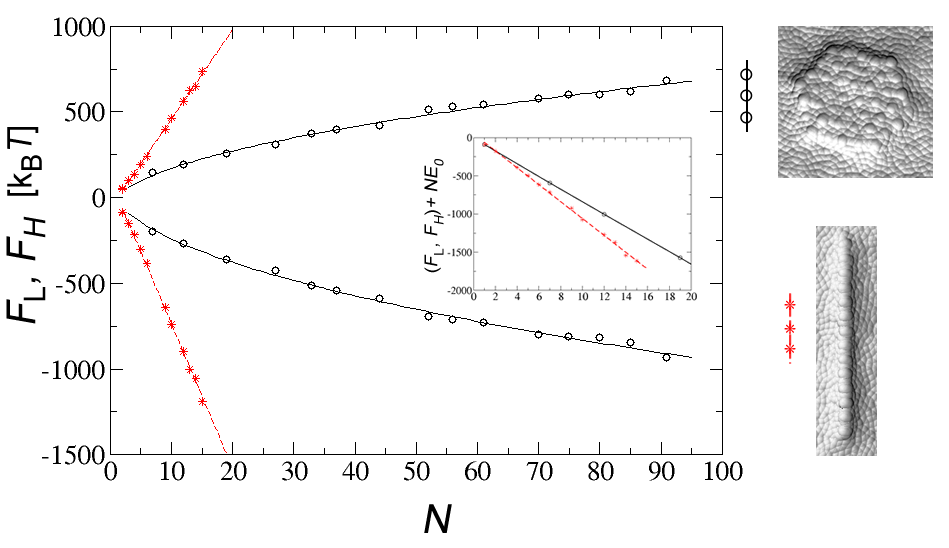}
\caption{Left panel: Difference in bending, $(F_L-F_H)_{\rm bend}$, and binding, $(F_L-F_H)_{\rm bind}$, energies between linear and hexagonal aggregates as a function of particle number $N$ at $\kappa_b=20k_{\rm{B}}T$, $D_0=10.9k_{\rm{B}}T$ and $\sigma_{np}=3\sigma$. The dashed line indicates the total energy difference between the two configurations. Right panel: typical membrane profiles underneath the aggregates in this regime.}
\label{fig3}
\end{figure}

 To understand the unexpected stability of the linear aggregates over the close-packed structures 
 in  the regime where linear aggregation occurs, we measured the energy of the system associated with linear and hexagonal aggregates. 
The l.h.s of Fig.~\ref{fig3} shows explicitly how the total energy difference between linear and hexagonal aggregates, computed for the same values of $\gamma, D_0$ and $k_b$, as a function of particle number $N$, is partitioned between the bending $(F_L-F_H)_{\rm bend}$ and the binding   $(F_L-F_H)_{\rm bind}$ contribution. This analysis reveals that despite the smaller bending cost, hexagonal aggregates  
provide a fairly small gain in binding energy when compared to linear aggregates, and this leads to a net energy balance that favors the latter.
It is worth mentioning that we monitored the difference in free energy due to the surface tension between the two 
configurations, and found it to be indeed negligible. We  also checked that linear aggregates form for our largest nanoparticles, $Z=6$.

To rationalize these numerical data we offer the following scaling argument.
A quick look at the typical surface deformations in this region of the phase diagram (see snapshots on the r.h.s of Fig.~\ref{fig3}) suggests that in either linear or hexagonal configuration the contribution to the system energy can be split in two parts. 
The first part comes from the overall deformation of the membrane due to the collective arrangement of the particles. The second part
comes from the shallow surface indentations  (corrugations) produced by each particle on top of the overall deformation.
Let's assume that the energy due to the corrugation is fairly independent of the overall arrangement of the aggregates.
 We can think of it as a particle self energy $e_0$ that is constant for a given $\kappa_B$, $\gamma$ and $D_0$. The total self-energy is than  $E_0=e_0N$. 

When particles arrange into linear structures ($L$), they generate a channel-like profile in the membrane with length proportional to the number of the nanoparticles $N$ and width proportional to $\sigma_{np}$. The bending energy of the channel can be estimated using the standard elastic energy $\frac{\kappa_B}{2}(A/R^2)$ \cite{landau} with $A$ being the area and $1/R$ being the constant curvature of the deformation. Ignoring the energy due to the contribution of the surface tension and subtracting the contribution of the particles' self energy, 
we can write the total free energy of the channel as $F_L-E_0\approx 2\pi\alpha(\frac{\kappa_B}{2}-D_0 \sigma^2_{np})N$, 
where $0<\alpha<1$ is a parameter that accounts for the degree of surface wrapping per nanoparticle, and is 
related to the overall height of the channel.
Close-packed hexagonal ($H$) arrangements form a flat, two-dimensional imprint  of lateral size proportional to $\sqrt{N}$. 
In this case, apart from a geometrical prefactor, the free energy due to the rim of the imprint scales as $F_H-E_0\approx \pi\alpha(\frac{\kappa_B}{2}(1+N^{-1})-D_0 \sigma^2_{np}) N^{\frac{1}{2}}$. In fact, here the area is proportional to the length of the rim and grows as $\sqrt{N}$, and the $N^{-1}$ term accounts for the small bending cost associated with the in-plane curvature of the rim $k_B/2c^2A$ with $c\sim N^{-1/2}$.
Structural stability requires both free energies to be negative ($D_0>\frac{\kappa_B}{2\sigma_{np}^2}$ for large $N$), which results in $F_L<F_H$, for sufficiently large values of $N$ making the linear aggregates more stable than the isotropic ones. 
In other words, the gain in binding energy overwhelms the larger cost in bending.
\vspace{5mm}
\vspace{25mm}

In conclusion, we have computed a phase diagram showing the different aggregates formed by  
nanoparticles adsorbing onto a lipid bilayer as a function of the surface bending rigidity and nanoparticles adhesive energy.
Our main result is that for a wide range of  bending rigidities $\kappa_b\approx 10-100 k_{\rm B}T$,
nanoparticles can  organize into linear aggregates $-$ provided the binding energy is sufficiently large.

Although linear aggregates are expected to form on elastic (polymerized) surfaces due to the global constraints 
imposed on the surface deformations by the stretching rigidity $K_s$ ( at least in the large $K_s$ limit)~\cite{cacciuto},
for fluid membranes $K_s=0$. Our result is therefore quite different than the expected, and usually assumed, 
isotropic aggregation mediated by  either local isotropic deformations of the surface or due to hydrophobic mismatch. 
The binding energy of the nanoparticles,  the missing ingredient in studies of aggregation of membrane inclusions, is the key to rationalize
this phenomenology. We hope our work will induce further analysis of membrane-mediated interactions between adhering nanoparticles.
\section*{ACKNOWLEDGMENTS}
This work was supported by the National Science Foundation under Career Grant No. DMR-0846426.


\begin{thebibliography}{99}
\bibitem{kostarelos} T. A.-J. Wafa​‌ and K. Kostarelos​‌, Nanomedicine 2, 85 (2007). 
\bibitem{costanzo} P. J. Costanzo, E. Liang, T. E. Patten, S. D. Collins, and R. L. Smith, Lab Chip 5, 606 (2005).
\bibitem{webb} R. J. Mart, K. P. Liem and S. J. Webb, Pharm. Res. 26, 1701 (2009).
\bibitem{safran} H. Aranda-Espinoza, A. Berman, N. Dan, P. Pincus and S. Safran, Biophys J. 71, 648 (1996).
\bibitem{pincus} M. Goulian, R. Bruisma and P. Pincus, Europhys. Lett. 22, 145 (1993).
\bibitem{netz} R.R. Netz, J. Phys. I France 7, 833 (1997).
\bibitem{fournier} P.G. Dommersnes and J.-B. Fournier, Eur. Phys. J. B, 12, 9 (1999).
\bibitem{chou} T. Chou, K. S. Kim and G. Oster, Biophysical Journal, 80, 10075 (2001).
\bibitem{deserno} C. Yolcu, I. Z. Rothstein, and M. Deserno, Europhys. Lett, 96, 20003 (2011).
\bibitem{chen} S. Mkrtchyan, C. Ing, and J. Z. Y. Chen, Phys. Rev. E 81, 011904 (2010).
\bibitem{killian} J. A. Killian, Biochim. Biophys. Acta Rev. Biomembr. 1376, 401 (1998).
\bibitem{leibler} S. Leibler, J. Phys. France 47, 507-516, (1986).
\bibitem{lipowsky} F. J\"{u}licher and R. Lipowsky, Phys. Rev. Lett. 70, 2964 (1992).
\bibitem{pep} R.N. Frese, J. C. Pamies, J. D. Olsen, S. Bahatyrova, C.D. Van Der Weij-De Wit, T. J. Aartsma, C. Otto, N. Hunter, D. Frenkel and R. Van Grondelley, Biophys. J. 94, 640 (2008). 
\bibitem{baumgart} T. Baumgart, S. T. Hess, W. W. Webb, Nature, 425, 821-824, (2003). 
\bibitem{safinya} I. Koltover, J. O. R\"{a}dler and C. R. Safinya, Phys. Rev. Lett. 82, 1991 (1999). 
\bibitem{safinya1} I. Koltover, T. Salditt and C. R. Safinya, Biophys J. 77, 915 (1999). 
\bibitem{ho} J.-S. Ho and A. Baumg\"{a}rtner, Europhys. Lett. 12, 295 (1990); A. Baumg\"{a}rtner and J.-S. Ho, Phys. Rev. A 41, 5747 (1990).
\bibitem{nelson} Y. Kantor, M. Kardar, and D. R. Nelson, Phys. Rev. Lett. 57, 791 (1986).
\bibitem{gompper} G. Gompper and D. M. Kroll, J. Phys.: Condens. Matter 9, 8795 (1997).
\bibitem{smit} M. Kranenburg, M. Venturoli and B. Smit, Phys. Rev. E 67, 060901 (2003).
\bibitem{li} H. Yuan1, C. Huang, J. Li, G. Lykotrafitis and Sulin Zhang, Phys. Rev. E 82, 011905 (2010).  
\bibitem{valleau} G. M. Torrie, J. P. Valleau, J. Comp. Phys. 23, 187 (1977)
\bibitem{weitz} L. Ramos, T. C. Lubensky, N. Dan, P. Nelson and D. A. Weitz, Science, 286, 2325 (1999).
\bibitem{landau}  L. D. Landau and E. M. Lifshitz, {\it Theory of Elasticity}, Pergamon: New York, (1970).
\bibitem{cacciuto} A. \v{S}ari\'c and A. Cacciuto, Soft Matter, 7, 8324 (2011).; A. \v{S}ari\'c and A. Cacciuto, Soft Matter, 7, 1784 (2011).

\end{thebibliography}
\end{document}